\documentclass{emulateapj}
\usepackage{psfig,epsfig}

\newcommand{\til}{\raisebox{-0.7ex}{\ensuremath{\tilde{\;}}}}

\newcommand{\mc}{\multicolumn}
\newcommand{\expnt}[2]{\ensuremath{#1 \times 10^{#2}}}   % scientific notation
\newcommand{\gsim}{\gtrsim}
\newcommand{\lsim}{\lesssim}

\newcommand{\hr}{\ensuremath{^{\rm h}}}
\newcommand{\mn}{\ensuremath{^{\rm m}}}

\newcommand{\xmm}{\textit{XMM-Newton}}
\newcommand{\asca}{\textit{ASCA}}
\newcommand{\axj}{AX~J183528\ensuremath{-}0737}
\newcommand{\mass}{2MASS~J18352582\ensuremath{-}0736501}
\newcommand{\sct}{Sct~X-1}
\newcommand{\ginga}{\textit{Ginga}}
\newcommand{\rxte}{\textit{RXTE}}
\newcommand{\heao}{\textit{HEAO 1}}
\newcommand{\ariel}{\textit{Ariel V}}
\slugcomment{Accepted by ApJ}

\begin{document}

\shorttitle{The Position and Counterpart of Scutum X-1}
\shortauthors{Kaplan et~al.}

\title{Lost and Found: A New Position and Infrared Counterpart
for the X-ray Binary Scutum X-1\altaffilmark{1,2}}

%% \author{David L.\ Kaplan\altaffilmark{3}, Alan M.~Levine, Deepto
%%   Chakrabarty, Edward H.~Morgan}
%% \affil{Kavli Institute for Astrophysics and Space Research,
%%   Massachusetts Institute of Technology, 77 Massachusetts Avenue,
%%   37-664H, 37-575, 37-626A, and 37-567, Cambridge, MA 02139} 
%% \email{dlk, aml, deepto, emh@space.mit.edu}
%% \author{D.~K.~Erb}
%% \affil{Harvard-Smithsonian Center for Astrophysics, MS 20, 60 Garden
%%   Street, Cambridge, MA 02138} 
%% \email{derb@cfa.harvard.edu}
%% \author{P.~B.~Cameron}
%% \affil{Division of Physics, Mathematics, and Astronomy, MS 105-24,
%%   California Institute of Technology, Pasadena, CA 91125.}
%% \email{pbc@astro.caltech.edu}
%% \author{\and\ D.-S.~Moon}
%% \affil{Department of Astronomy and Astrophysics,
%% University of Toronto,
%% 60 St. George Street, Toronto, ON  M5S 3H8
%% Canada}
%% \email{moon@astro.utoronto.ca}

\author{David L.\ Kaplan\altaffilmark{3,4,5}, Alan M.~Levine\altaffilmark{4}, Deepto
   Chakrabarty\altaffilmark{4,5}, Edward H.~Morgan\altaffilmark{4},
   Dawn~K.~Erb\altaffilmark{6},  B.~M.~Gaensler\altaffilmark{7},
   Dae-Sik~Moon\altaffilmark{8}, \&\
   P.~Brian~Cameron\altaffilmark{9}}

\altaffiltext{1}{This paper includes data gathered with the 6.5 meter
Magellan Telescopes located at Las Campanas Observatory, Chile.}
\altaffiltext{2}{Partially based on data obtained at the W.~M.~Keck
  Observatory, 
  which is operated as a scientific partnership among the California
  Institute of Technology, the University of California, and NASA, and
  was made possible by the generous financial support of the
  W.~M.~Keck Foundation.}
\altaffiltext{3}{Pappalardo Fellow}
\altaffiltext{4}{Kavli Institute for Astrophysics and Space Research,
   Massachusetts Institute of Technology, Cambridge, MA 02139,
   {dlk, aml, deepto, ehm@space.mit.edu}} 
\altaffiltext{5}{Also Department of Physics, Massachusetts Institute
  of Technology, Cambridge, MA 02139} 
\altaffiltext{6}{Harvard-Smithsonian Center for Astrophysics,
    MS 20, 60 Garden Street, Cambridge, MA 02138,
    {derb@cfa.harvard.edu} }  
\altaffiltext{7}{School of Physics A29, The University of Sydney, NSW
  2006, Australia, {bgaensler@usyd.edu.au}} 
\altaffiltext{8}{Department of Astronomy and Astrophysics,
 University of Toronto, 50 St. George Street, Toronto, ON  M5S 3H8
 Canada, {moon@astro.utoronto.ca}} 
\altaffiltext{9}{Division of Physics, Mathematics, and
    Astronomy, MS 105-24,    California Institute of Technology,
    Pasadena, CA 91125, {pbc@astro.caltech.edu}}

\begin{abstract}
Using archival X-ray data, we find that the catalog location of the
X-ray binary Scutum~X-1 (\sct) is incorrect, and that the correct
location is that of the X-ray source \axj, which is $15\arcmin$ to the
west.  Our identification is made on the basis of the 112-s pulse
period for this object detected in an \xmm\ observation, as well as
spatial coincidence between \axj\ and previous X-ray observations.
Based on the \xmm\ data and archival \rxte\ data, we confirm secular
spin-down  over 17~years with period derivative
$\dot P\approx \expnt{3.9}{-9}\mbox{ s s}^{-1}$, but do not detect a
previously reported X-ray iron fluorescence line.  We identify a
bright ($K_{\rm s}=6.55$), red ($J-K_{\rm s}=5.51$), optical and
infrared counterpart to \axj\ from 2MASS, a number of mid-IR surveys,
and deep optical observations, which we use to constrain the
extinction to and distance of \sct.  From these data, as well as
limited near-IR spectroscopy, we conclude that \sct\ is most likely a
binary system comprised of a late-type giant or supergiant and a
neutron star.
\end{abstract}


\section{Introduction}
\label{sec:int}
By scanning the Galactic plane with an X-ray payload on a sounding
rocket, \citet{hbg+74} discovered an unusual X-ray source in the
constellation Scutum, which they named \object[X Sct X-1]{Scutum~X-1}
(hereafter \sct).  \sct\ was subsequently detected using X-ray
instruments on the \textit{Copernicus}, \ariel, \heao, and \ginga\
satellites (\citealt*{cmd75}; \citealt{mbh+79,rjb+80,cll+84,kktt91,yk93}); in
contrast, it was not detected, at least unambiguously, in surveys done
using the \textit{Uhuru} and \textit{EXOSAT} satellites \citep[Fourth
Uhuru Catalog]{fjc+78,wnt+88}.  In summary, these observations
determined that (i) the X-ray flux varies over a range as wide as
$<0.3$~mCrab to 20~mCrab, and this variability occurs on a variety of
timescales \citep{cmd75,mbh+79,rjb+80,cll+84,wnt+88,kktt91,yk93}; (ii)
the emission has a hard spectrum with a significant low-energy cutoff
that implies an interstellar absorption column density of $\gsim
10^{23}\mbox{ cm}^{-2}$; (iii) the source is an X-ray pulsar with a
111-s pulsation period \citep{makino88}; and (iv) the X-ray spectrum
may show line emission at 6.4~keV, presumably from iron K fluorescence
\citep{kktt91}.  Based on its similarity to other sources at the same
Galactic longitude, \citet{kkk+90} concluded that \sct\ is likely an
X-ray binary in the Scutum arm of the Galaxy at about 10~kpc distance.

%Aside from identifying it as an X-ray binary, however, the previous
%observations have made few conclusion about what \sct\ is: what sort
%of compact object, companion star, orbit, and method of accretion
%(disk or wind).  Poor spatial resolution and high background from the
%Galactic ridge made precise localization difficult, and this inhibited
%followup measurements.

The most accurate information on the celestial location of \sct\ comes
from \citet{rjb+80}, who presented positions of X-ray sources measured
with the modulation collimator (MC) on the \heao\ satellite (\heao\
A3).  They derived a grid of diamond-shaped error boxes, but concluded
that only two of the diamonds were consistent with  earlier
observations \citep{hbg+74,cmd75,mbh+79}, viz., those centered at
$18\hr34\mn49\fs54$, $-07\degr37\arcmin51\farcs3$ and
$18\hr33\mn46\fs48$, $-07\degr38\arcmin40\farcs0$ (both B1950; the
corresponding J2000 positions are $18\hr37\mn32\fs14$,
$-07\degr35\arcmin13\farcs7$ and $18\hr36\mn29\fs10$,
$-07\degr36\arcmin06\farcs9$).  Each error diamond was approximately $\pm
14\arcsec$ in Right Ascension by $\pm 76\arcsec$ in Declination.

In this paper we present the results of our analysis of archival X-ray
observations of \sct\ with \asca, \xmm, and \rxte\footnote{After
submission of this manuscript, we were made aware of a similar
analysis performed by another group (see \citealt{hgh04} and
\url{http://www.astro.columbia.edu/{\til}jules/axj1835.4-0737/}).
Their conclusions are largely the same as our own.}. We find that the
source is \textit{not} in either of the two error boxes chosen by
\citet{rjb+80}, but is instead in the next box to the west.  The
presence of a 112-s period pulsation (having evolved from the 111-s
period that was found previously) and the coincidence in position with
one of the \heao\ error boxes make us confident of our identification.
We describe the X-ray analysis in \S~\ref{sec:xray}, and identify a
near-IR counterpart from the Two-Micron All-Sky Survey (2MASS;
\citealt{2mass}) and mid-IR surveys in \S~\ref{sec:ir}.  Finally, we
give our discussion  in \S~\ref{sec:disc} and conclusions in \S~\ref{sec:conc}.

\begin{figure}
\psfig{file=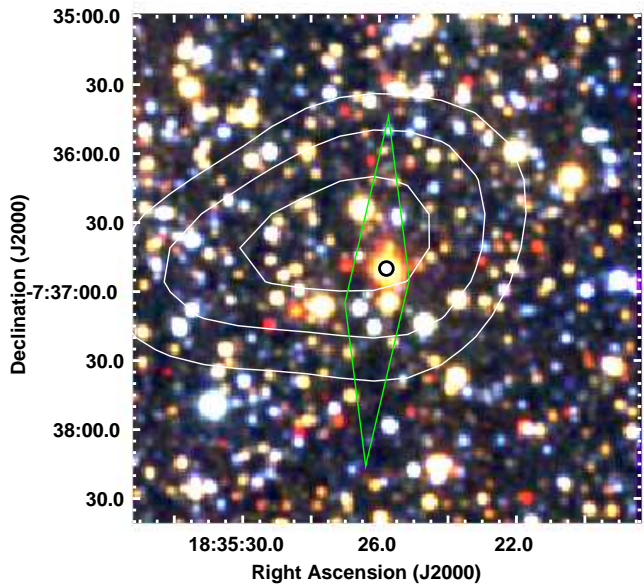,width=0.48\textwidth,clip=}
\caption{Our identification of \axj\ with the X-ray binary \sct\ and
  the bright ($K_s=6.5$) near-IR counterpart \mass.  The image is a
  red-green-blue
 composite of 2MASS $K_{\rm s}$, $H$, and $J$.  The contours show
  smoothed \asca\ GIS emission, the \xmm\ position is shown by the
  circle with a radius of $3\arcsec$ (expanded from a $1\arcsec$ radius
  for visibility), and we also show one of the \heao\ position
  diamonds, from \citet{rjb+80}.
\label{fig:image}}
\end{figure}

\section{Archival X-ray Data Analysis}
\label{sec:xray}
We found that the X-ray source \object[AX J1834.4-0737]{\axj},
discovered by \citet{smk+01} in their survey of the Galactic plane
with \asca\ (this specific observation is from 1997~October~13), was close to but not consistent with the published
positions of \sct\ \citep{rjb+80}.  However, there were too few counts
in the \asca\ data to perform a reasonably sensitive search for
periodicities and thereby uniquely identify \axj: \citet{smk+01}
measured a 0.7--7~keV count rate of $26.9\mbox{ ksec}^{-1}\mbox{
GIS}^{-1}$ in an observation with 4~ksec of good exposure, so there
are only $\approx 220$ counts (summed over the two GIS detectors).

A 17-ks \xmm\ observation of \axj\ (observation 0203850201) was performed
on 2004~September~18/MJD~53266.26.  There is one bright  source
detected in the EPIC-pn and EPIC-MOS images (the source is not
detected in the RGS data). This source, which is located at (J2000)
$18\hr35\mn25\fs8$, $-07\degr36\arcmin50\arcsec$ with statistical
position uncertainties of $0\farcs1$ (and absolute uncertainties of
$\lsim 1\arcsec$), is fully consistent with the position of \axj\ (see
Fig.~\ref{fig:image}).  It has a background-subtracted EPIC-pn count
rate of $0.390(9)\mbox{ s}^{-1}$ in the 0.5--8~keV band, and is
consistent with a point source given the angular resolution of \xmm.
Figure~\ref{fig:image} also shows that the positions of \axj\ determined
from both the \asca\ and \xmm\ observations are fully consistent with
one of the error boxes of \sct\ that was derived from the much earlier
\heao\ observations although it does not happen to be one of the two
error boxes selected as prime candidates for the source location by
\citet{rjb+80}.

\begin{figure}
\psfig{file=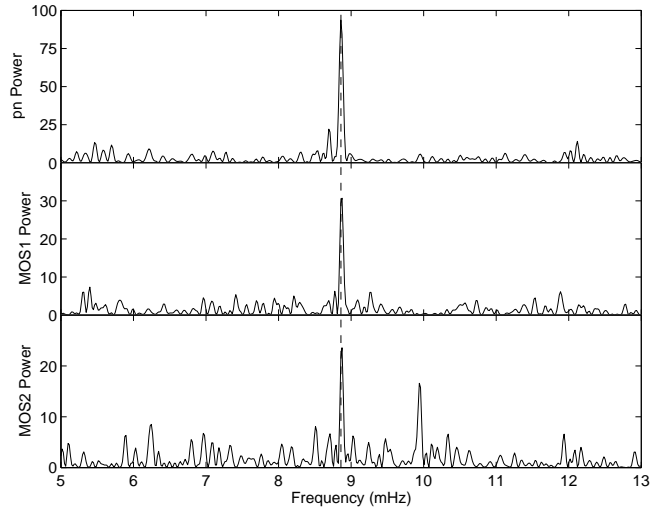,width=0.48\textwidth}
\caption{Power-spectra of \sct.  We show $Z_1^2$ power spectra for
  EPIC-pn (top), EPIC-MOS1 (middle), and EPIC-MOS2 (bottom) data in
  the 2--8~keV band.  The best-fit
  frequency from the EPIC-pn data (8.860(6)~mHz) is plotted as the
  vertical dashed line in all three panels.}
\label{fig:z12}
\end{figure}

\begin{figure}[h]
\psfig{file=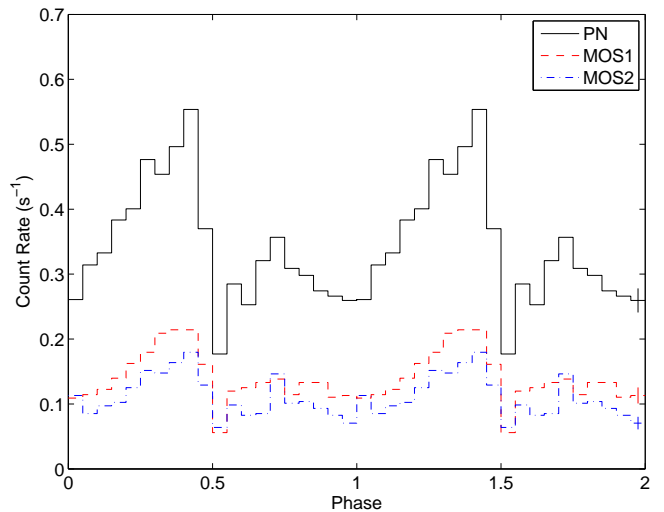,width=0.48\textwidth}
\caption{Pulse profiles of \sct\ in the 0.5--8~keV band, repeated
  twice for clarity.  We show two cycles of the events detected in the
  EPIC-pn (black solid line), EPIC-MOS1 (red dashed line) and
  EPIC-MOS2 (blue dash-dotted line) folded on the best-fit period of
  112.86~s; representative error bars are shown on the right-most
  points.}
\label{fig:pulse}
\end{figure}

\begin{deluxetable*}{l c c c c c c}
\tablecaption{\rxte\ Observation Summary\label{tab:xte}}
%\tabletypesize{\scriptsize}
\tablewidth{0pt}
\tablehead{
\colhead{ObsID} & \colhead{Epoch} & \colhead{Exposure} &
\colhead{Trans.\tablenotemark{a}} & \colhead{CR\tablenotemark{b}} &
\colhead{Ampl.\tablenotemark{c}} &\colhead{Signif.\tablenotemark{d}}\\
 & \colhead{(MJD)} & \colhead{(sec)} & \colhead{(\%)} & \colhead{(s$^{-1}$)}& \colhead{(s$^{-1}$)}\\
}
\startdata
20143-08-01-00  &50619.835381(14)&10361 &74  &81.98  &9.5(2)  & 222\\
20143-08-01-01  &50620.757608(13)&11579 &74  &82.55  &7.5(2)  & 259\\
20142-02-01-00  &50688.532073(30)&\phn7840  &68  &59.44  &1.6(2)  &  \phn46\\
20142-02-03-00  &50765.147085(46)&\phn8148  &70  &71.82  &0.7(2)  &  \phn19\\
\enddata
\tablenotetext{a}{Average collimator transmission of \sct.}
\tablenotetext{b}{Mean count rate including background.}
\tablenotetext{c}{Amplitude of fitted sinusoid.}
\tablenotetext{d}{Significance of period detection, in units of $\sigma$.}
\end{deluxetable*}

\subsection{Timing Analysis}

To see whether or not this source is \sct, we searched for evidence of
the 111-s period detected by \citet{makino88} and \citet{kktt91}.  We
extracted events (using XMMSAS release 20050815) from the pn and MOS
data with \texttt{PATTERN}~$\leq 4$ (singles and doubles) and energies
between 2 and 8~keV within a radius of 500~pixels ($25\arcsec$).  We
barycentered the event arrival times and constructed $Z_1^2$ power
spectra \citep{bbb+83}.  Searching the $Z_1^2$ power spectrum of the
EPIC-pn data for periods from 75--210~s (all much longer than the
frame time of 73~ms), we find a very significant peak ($Z_1^2$
amplitude of 93.6, which gives a probability of $\expnt{2}{-41}$ in a
single trial) at $8.860\pm 0.006$~mHz ($112.86\pm 0.08$~s; see
Fig.~\ref{fig:z12}), and the power spectra made from the MOS data also
have peaks at $112.69\pm 0.16$~s (MOS1) and $112.69\pm 0.19$~s (MOS2),
where the uncertainties have been calculated according to
\citet{ransom01}.  There is a small peak at a $Z_1^2$ power of 22 in
the EPIC-pn power spectrum, but this is likely a sidelobe of the main
peak.  No other strong peaks were found in the EPIC-pn or MOS1
spectra, while in the MOS2 spectrum there is a peak at a period of
100.6~s which reaches a power about $1/2$ that of the peak at
$112.69$~s.  However, given the much lower amplitude of the main MOS2
peak (23.6, compared to 93.6 for the pn) the secondary peak is not very
strong overall (single trial probability of $\sim 10^{-7}$).  Based on
this, we conclude that a $\sim112$~s pulse period was detected, and
that the 100.6~s period is likely either a noise spike or an
instrumental effect in the MOS2 data.  Folding the data on the 112-s
period, we see (Fig.~\ref{fig:pulse}) strong, asymmetric pulsations
with a pronounced interpulse, not unlike the pulse profiles shown in
\citet{kktt91}.  The pn data have an rms pulsed fraction of 29\%.

\begin{figure}[b]
\psfig{file=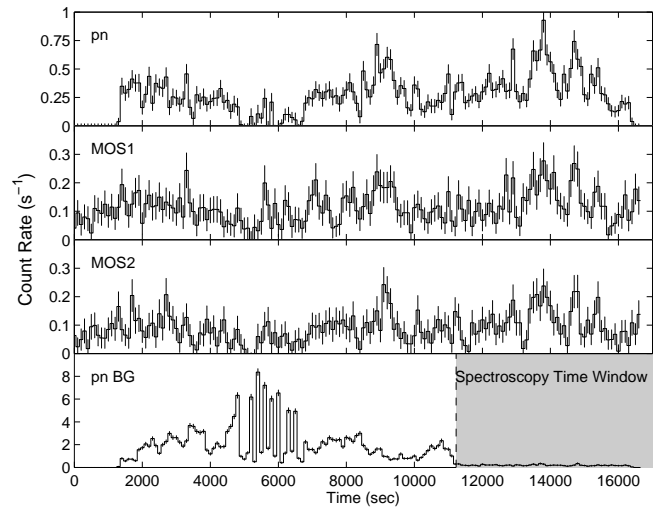,width=0.48\textwidth}
\caption{Background-subtracted lightcurves as a function of time since
  the start of the observation (0.5--8~keV band, with
  200-s bins) made from EPIC-pn, EPIC-MOS1, and EPIC-MOS2 data, as
  labeled.  We also show the EPIC-pn background (over all energies)
  lightcurve in the bottom panel; this background lightcurve was
  scaled down by a factor of 25 before subtraction from the EPIC-pn
  lightcurve (the EPIC-MOS background lightcurves were similar).  The
  bright flaring near 6000~s was responsible for the decrease in the
  source count-rate at the same times due to telemetry saturation.
  For the spectroscopy (\S~\ref{sec:spec}) we only used the low
  background data from the shaded interval.}
\label{fig:lc}
\end{figure}

To examine the variability of the X-ray source during the \xmm\ exposure
and to search for
background flares, we binned the EPIC event data in 200-s bins to
construct lightcurves, which we show in Figure~\ref{fig:lc}.  In all
three EPIC detectors the background is high with episodes of strong
flaring for the first $2/3$ of the observation.  This flaring is
likely responsible (through telemetry saturation) for the dips in the
source lightcurves at $\approx 5000$~s, as the unfiltered EPIC-pn
count rate exceeds the limit of $400\mbox{ s}^{-1}$ and therefore the
cameras entered ``counting mode''\footnote{See
\url{http://xmm.vilspa.esa.es/external/xmm\_user\_support/documentation/uhb/node95.html}.}
(note that excluding the saturated regions has no effect on the
period determination).  However, even away from the times of strong
flaring in the background, the source strength appears to vary by a
factor of $\approx 2$; see the small increases near 9000~s and
especially the variations in count rate after the background dies out
near 11000~s.  These variations are similar to those seen by
\citet{kktt91}.  We also find that the spectral hardness stays
relatively constant while the overall count rate varies.

\begin{deluxetable}{l l  l}
\tablecaption{Period measurements for \sct\label{tab:per}}
\tablewidth{0pt}
\tablehead{
\colhead{MJD} & \colhead{Instrument} & \colhead{Period} \\
\colhead{(UT)} & &\colhead{(sec)} \\
}
\startdata
47479.6 & \ginga/LAC & 111.001(4) \\
48142.7 & \ginga/LAC & 111.194(3) \\
%50619.9 & \rxte/PCA & 112.228(4) \\
50619.8 & \rxte/PCA & 112.23(4) \\
50620.8 & \rxte/PCA & 112.22(4) \\
%50684.4 & \rxte/PCA & 112.29(20)  \\
50688.5 & \rxte/PCA& 112.28(6)  \\
%50727.3 & \rxte/PCA& 112.26(8)  \\
50765.1 & \rxte/PCA& 112.37(7)  \\
53266.4 & \xmm/EPIC-pn & 112.86(8)  \\
53266.4 & \xmm/EPIC-MOS1 & 112.69(16)  \\
53266.4 & \xmm/EPIC-MOS2 & 112.69(19)  \\
\enddata
\tablecomments{Quantities in parentheses are 1$\,\sigma$ uncertainties
  on the last digit.  \ginga/LAC data are rom \citet{yk93}.} 
\end{deluxetable}

\sct\ has also been observed on a few occasions by the \textit{Rossi
X-ray Timing Explorer} (\rxte) satellite. We have analyzed portions of
the observations in which the Proportional Counter Array (PCA) was
steadily pointed at the target location; other portions of the
observations included short pointed observations or scans over regions
around the target location in vain attempts to precisely determine the
position of the source.  The longer pointed observations comprise data
from two observations in 1997~June, one in 1997~August, and one in
1997~November, and have exposure times ranging from 4 to 12~ks (see
Tab.~\ref{tab:xte}).  Data from all 5 proportional counter units
(PCUs) in the energy range 4--10~keV were binned into 1~s time
bins. The resulting light curves were fit with a sinusoid plus 3
harmonics over trial periods between 110 and 115~s. In all the
observations a clear dip in $\chi^2$ is seen near 112.3~s, although
the amplitude of the fitted sinusoid varied over a wide range. For
each of the four observations, the error on the period was determined
from the range of trial periods with $\chi^2 < \chi_{min}^2 + \Delta
\chi^2$ where $\Delta \chi^2$ was the full range of variation of
$\chi^2$ found in a comparable region of period space with no  signal.  We
list the final period measurements in Table~\ref{tab:per}.

\subsection{Spectral Analysis}
\label{sec:spec}

\setlength{\tabcolsep}{2pt}
\begin{deluxetable}{l l l l l}
\tablecaption{Spectral fits to EPIC data of \sct\label{tab:spec}}
\tablewidth{0pt}
\tablehead{
\colhead{Parameter} & \mc{4}{c}{\dotfill Fit Type\tablenotemark{a}\dotfill} \\ 
 & \colhead{PL} & \colhead{TB} & \colhead{PL+Fe} & \colhead{PL+Fe} \\
}
\startdata
$N_{\rm H}$ ($10^{22}\mbox{ cm}^{-2}$) & 8.1(6) & 7.8(5)  &8.1 & 8.1(6)\\
$\Gamma$/$kT$ (keV) & 1.48(13) & $30_{-10}^{+28}$ &1.48 & 1.49(14)\\
Norm\tablenotemark{b} & 1.6(3) & 2.19(8) & 1.6 & 1.6(4)\\ 
EW$_{\rm Fe}$ (eV)\tablenotemark{c} & \nodata & \nodata & 20(38) & 23(40)\\
$\chi^2/$DOF & 84.5/118 & 83.6/118 & 84.2/120 & 84.2/117\\

$F_{\rm X}$ $(10^{-11}\mbox{erg s}^{-1}\mbox{
  cm}^{-2})$\tablenotemark{d} & 1.2 & 1.2 & 1.3 & 1.3 \\
\enddata
\tablecomments{Quantities in parentheses on the fit parameters
  (excluding the flux) are
  1$\,\sigma$ uncertainties 
  in the last digit;
  quantities without uncertainties were held fixed for the fit.  }
\tablenotetext{a}{Fit
  types are absorbed power-law with (PL+Fe) and without (PL) an iron
  line at 6.4~keV and thermal bremsstrahlung (TB).  In the first PL+Fe
  fit, the PL parameters are held 
fixed at the best-fit values from the PL fit.} 
\tablenotetext{b}{Either power-law normalization, in units of
  $10^{-3}\mbox{photons s}^{-1}\mbox{ 
  cm}^{-2}\mbox{ keV}^{-1}$ at 1~keV, or thermal bremsstrahlung
  normalization, in units of $\expnt{3.02}{-18}\int dV
  n_{e} n_{I}/(4\pi d^2) $, where $d$ is the distance to the source, $n_{e}$ and
  $n_{I}$ are the electron and ion number densities, and the integral
  is over the source volume (based on the \texttt{XSpec}
  \texttt{bremss} model).} 
\tablenotetext{c}{Equivalent width of a putative unresolved Fe line at 6.4~keV.}
\tablenotetext{d}{Model flux in the 0.5--10~keV band, corrected for absorption.}
\end{deluxetable}

We fit spectra to the EPIC data.  After filtering for single and
double events (as above), we excised the data taken when the
background was high, i.e., prior to 11~ks from the start of the
observation (Fig.~\ref{fig:lc}) and were
left with exposure times of 4.5, 5.3, and 5.3 ksec for the EPIC-pn,
MOS1, and MOS2 respectively, out of initial exposures of 16~ksec for
pn and 17~ksec for the MOSs.

We extracted source events from the same 500-pixel radii regions
that we used for timing.  For the background, we selected pn events
from a circle 2500~pixels ($125\arcsec$) in radius located near but
not containing \axj, as the source was close to a CCD boundary and we
could not define a proper background annulus that would be on the same
CCD.  For the MOS data, where \axj\ was located in the middle of a
CCD, we used a background annulus extending from 500 to 5000~pixels
($250\arcsec$) in radius.  We binned the data so that each bin has
$\geq 25$~counts, and then we generated ancillary response files
(ARFs) and redistribution matrix files (RMFs) for each instrument and
fit the data in \texttt{Sherpa}\footnote{Part of the Chandra Interactive Analysis of Observations (CIAO), \url{http://cxc.harvard.edu/ciao/}.}

\begin{figure}
\psfig{file=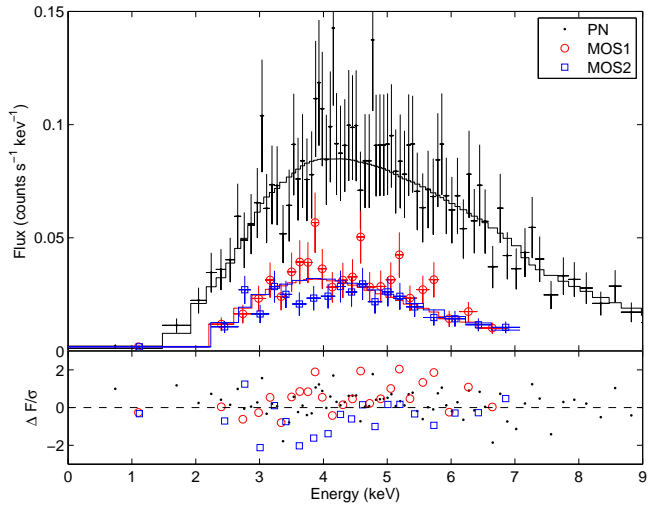,width=0.48\textwidth}
\caption{X-ray spectrum of \sct, from the XMM EPIC data.  The upper
  panel shows the best power-law fit to the data from EPIC-pn (black
  points), EPIC-MOS1 (red circles) and EPIC-MOS2 (blue squares).  The
  fit parameters are given in Table~\ref{tab:spec}.  The lower panel
  shows  residuals in units of $\sigma$.}
\label{fig:spec}
\end{figure}

The data from all three instruments are jointly well-fit by an
absorbed power-law; we give basic fit parameters in
Table~\ref{tab:spec} and show the fit in
Figure~\ref{fig:spec}. Attenuation factors were calculated using the
\texttt{phabs} model (which uses cross sections from \citealt{bm92}
and solar abundances).  The quality of the fit is good, as indicated
by a low reduced $\chi^2$ (0.7), and the residuals do not show any
significant systematic deviations.  The data could also be fit with a
very hot ($kT\approx 30$~keV) thermal bremsstrahlung model, but the
implied absorption was quite similar and at these temperatures the
thermal model is essentially a power-law\footnote{More complicated
models of thermal X-ray emission from optically thin hot plasmas, such
as those of \citet{rs77}, give similar results if the abundances of
metals are low.}.

Interestingly, \citet{kktt91} found evidence in the spectrum of \sct\
for an iron emission line at 6.4~keV with an equivalent width of
0.2--0.3~keV.  While not required to achieve a good fit to our data,
we examined whether the addition of such a line to the spectral model
would improve the fit.  To test this, we fit the EPIC data with two
models: the power-law model frozen to its best-fit values with the
addition of a line at 6.4~keV (third column in Tab.~\ref{tab:spec}),
and an absorbed power-law plus iron line where all parameters are free
(fourth column in Tab.~\ref{tab:spec}). The iron line was assumed to
be unresolved for both fits.  In neither case did the addition of the
iron line improve the fit significantly.  There was a small reduction
in $\chi^2$ from 84.5 to 84.2, but this is not significant.  The
best-fit values of the power-law parameters in the fourth fit did not
change appreciably from those in the first fit.  Overall, we find no
evidence for a line at 6.4~keV, and can set a 90\%-confidence upper
limit to the equivalent width of 96~eV.  This is formally inconsistent
with the results of \citet{kktt91}.  They discussed the difficulty in
properly subtracting the strong Galactic ridge emission, and this
could be the cause of the discrepancy, although the source could also
be variable as we discuss below.

\begin{figure*}
\plottwo{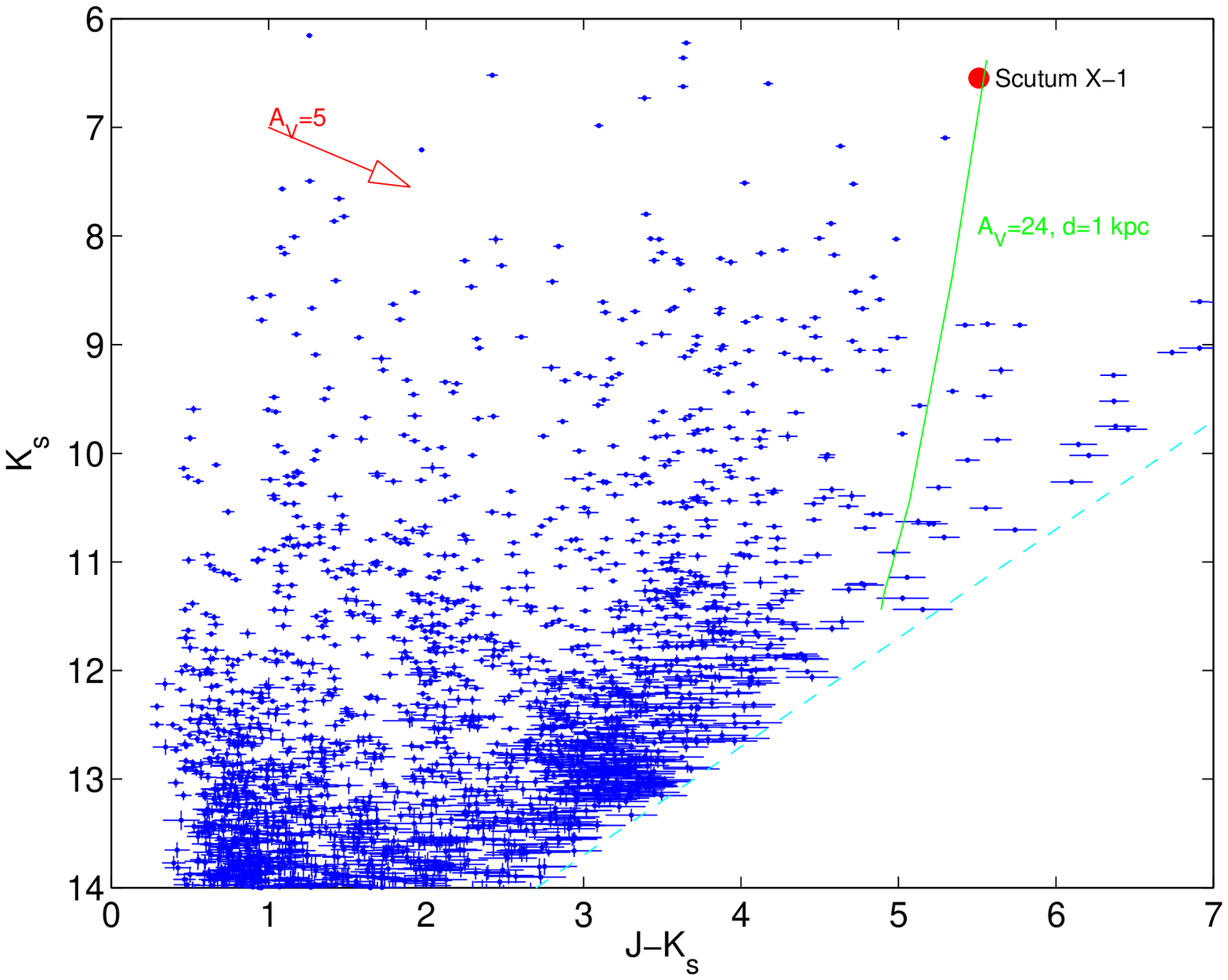}{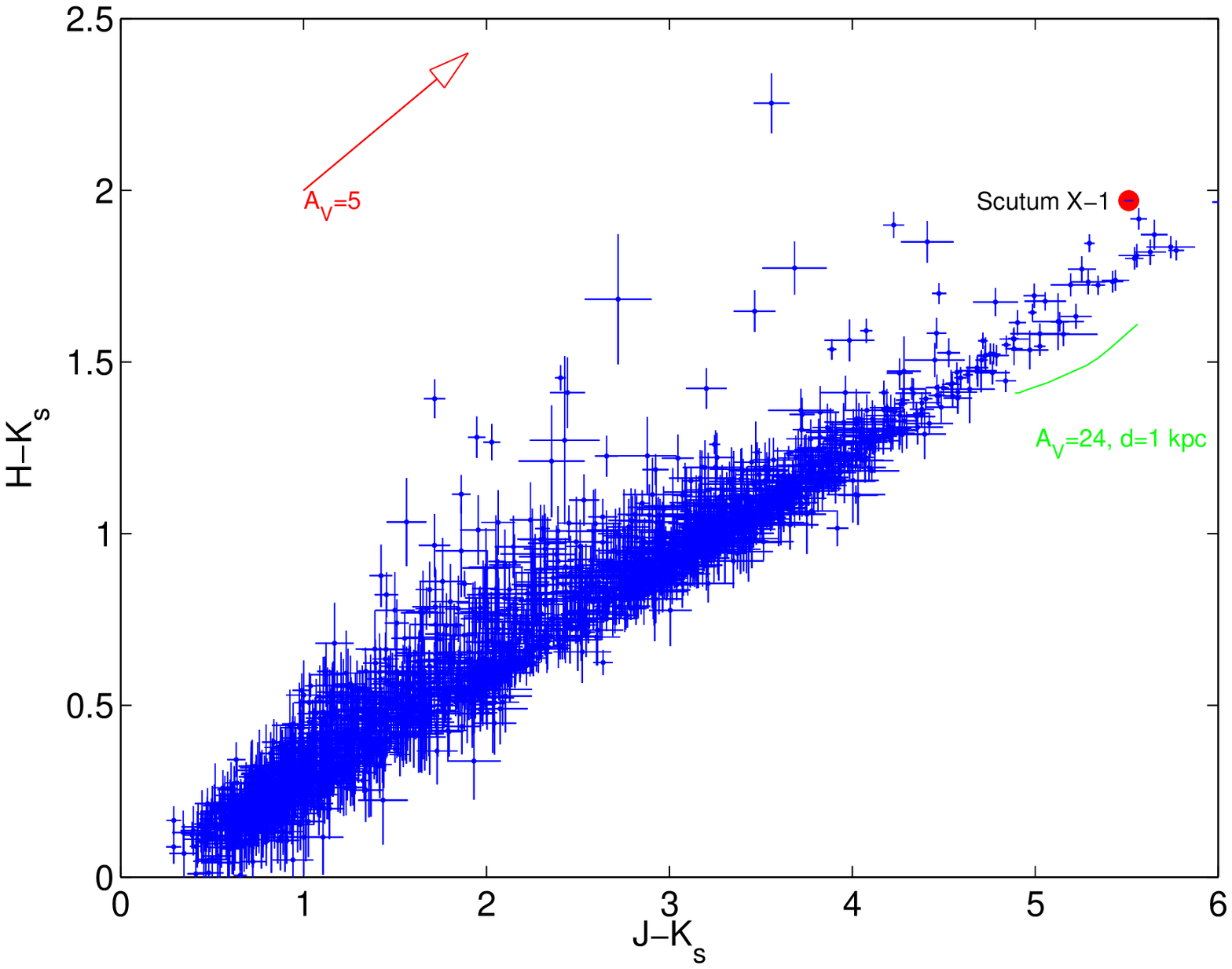}
\caption{Near-IR color-magnitude diagram (\textit{left}) and
  color-color diagram (\textit{right}) of the $\approx 2000$ stars
  within $500\arcsec$ of \mass, the proposed infrared counterpart to \sct, from 2MASS.  \mass\ is the red
  circle, the arrows indicates $A_V=5$, and we also plot a giant
  sequence (luminosity class III) at an extinction $A_V=24$ and a
  distance of 1~kpc, taken from \citet{allen}; this shows the stellar
  colors that imply the minimum extinction (\S~\ref{sec:oir}).  The
  empty region at the lower right of the color-magnitude diagram is
  from the 2MASS sensitivity limit, as indicated by the dashed line.}
\label{fig:cmd}
\end{figure*}

\section{Optical/Infrared Counterpart}
\label{sec:ir}

\subsection{Archival Data}
We found a potential IR counterpart to \axj\ in the 2MASS Point Source
Catalog: \mass, which is $<0\farcs2$ away from the \xmm\ source (see
Fig.~\ref{fig:image}).  This object is rather bright and red, with
$K_{\rm s}=6.55\pm0.02$, $J-K_{\rm s}=5.51\pm0.03$ and $H-K_{\rm
s}=1.97\pm0.03$.  It is among the brightest of the sources in the
2MASS color-magnitude diagram of stars within $500\arcsec$
(Fig.~\ref{fig:cmd}), and is considerably redder than most sources of
similar magnitudes.  Just on the basis of the magnitude, the
association with \sct\ is very probable, as we find $\approx 10^{-5}$
sources in this region with $K_{\rm s}<6.6$ per square arcsecond, so
the chance for a random alignment within $0\farcs2$ is a negligible
$\expnt{2}{-6}$.

We also find \mass\ in other near- and mid-infrared catalogs: the Deep
Near Infrared Survey of the Southern Sky (DENIS; \citealt{edd+99})
database, the \textit{Spitzer Space Telescope} Galactic Legacy
Infrared Mid-Plane Survey Extraordinaire \citep[GLIMPSE; ][]{bcb+03}
data-base, the \textit{Midcourse Space Experiment} \citep[{\em
MSX};][]{pec+01}\footnote{See
\url{http://irsa.ipac.caltech.edu/Missions/msx.html}.} Point Source
Catalog (MSX6C; \citealt{epk+03}), and the \textit{IRAS} Point Source
Catalog (PSC).  We summarize all of the infrared data on \mass\ in
Table~\ref{tab:ir}.

In DENIS, the source is DENIS~J183525.8$-$073650, and it is detected
with $J=11.82\pm0.06$, $K_{\rm s}=6.33\pm0.10$, but is not detected in
$i$-band (where the limiting magnitude is $\approx 18.5$).  These
magnitudes are slightly different from the 2MASS values, even allowing
for color transformations \citep{c01}, but we note that the $K_{\rm
s}$-band measurement is near the DENIS saturation limit.

In the GLIMPSE Archive (which is more complete but less reliable than
the GLIMPSE Catalog), it is listed as SSTGLMA~G024.3361+00.0657, with
fluxes of $3.49\pm0.13$, $2.61\pm0.10$, $4.17\pm0.12$, and
$3.42\pm0.20$~Jy at 3.6, 4.5, 5.8, and 8.0~$\mu$m, respectively.  The
variation of the fluxes as a function of wavelength is puzzling, since
a star would generally have monotonically decreasing values.  This is
likely due to non-linearity/partial saturation of this source, and
indeed we note that SSTGLMA~G024.3361+00.0657 is {not} included in the
GLIMPSE Catalog, the data-quality flag indicates that no non-linearity
correction was applied (bit 19), and the fluxes are above the nominal,
albeit conservative\footnote{See the GLIMPSE Data Products Document
at\\
\url{http://data.spitzer.caltech.edu/popular/glimpse/20050415\_enhanced\_v1/Documents/glimpse\_dataprod\_v1.5.pdf}.},
saturation limits of 0.44, 0.45, 2.9, and 1.6 Jy in all bands; the
radial profiles in the 3.6- and 4.5-$\mu$m images also show effects of
saturation.  Finally, in MSX6C, it is source MSX6C~G024.3359+00.0656,
and in the \textit{IRAS} PSC it is IRAS~18327$-$0739.

\begin{deluxetable}{c c c c}
\tablecaption{Optical and Infrared Fluxes of \mass\label{tab:ir}}
\tabletypesize{\scriptsize}
\tablewidth{0pt}
\tablehead{
\colhead{Catalog/} & \colhead{Band} & \colhead{Wavelength} & \colhead{Flux} \\
\colhead{Instrument} & & \colhead{($\mu$m)} & \colhead{(Jy)} \\
}
\startdata
MagIC & $r^{\prime}$ & \phn0.6 & $<\expnt{3}{-7}$ \\
      & $i^{\prime}$ & \phn0.8 & $\expnt{1.3(2)}{-6}$ \\
DENIS & $J$ & \phn1.2 & 0.030(2) \\
 & $K_{\rm s}$& \phn2.2 & 2.0(2) \\
2MASS & $J$& \phn1.2 & 0.024(1) \\
 & $H$ &\phn1.7 & 0.40(1) \\
 & $K_{\rm s}$ &\phn2.2 & 1.60(4)\\ 
GLIMPSE & &\phn3.6 & 3.49(13)\\
 && \phn4.5 & 2.61(10) \\
 && \phn5.8 &  4.17(12) \\
 && \phn8.0 & 3.42(20) \\
MSX6C & A& \phn8.3 &  5.1\\
 & C& 12.1  & 7.8\\
 &D &14.7 & 6.5\\
 &E &21.3 & 3.9\\
IRAS & &12\phantom{.}\phn & 10.3 \\
\enddata
\tablecomments{Values in parentheses
  are  1$\,\sigma$ statistical uncertainties on the last digits, if
  available, but note that the 
  GLIMPSE data, especially from the 3.6- and 4.5-$\mu$m bands, are
  likely somewhat saturated.  See \S~\ref{sec:ir} for additional discussion.}
\end{deluxetable}

\begin{figure*}
\plotone{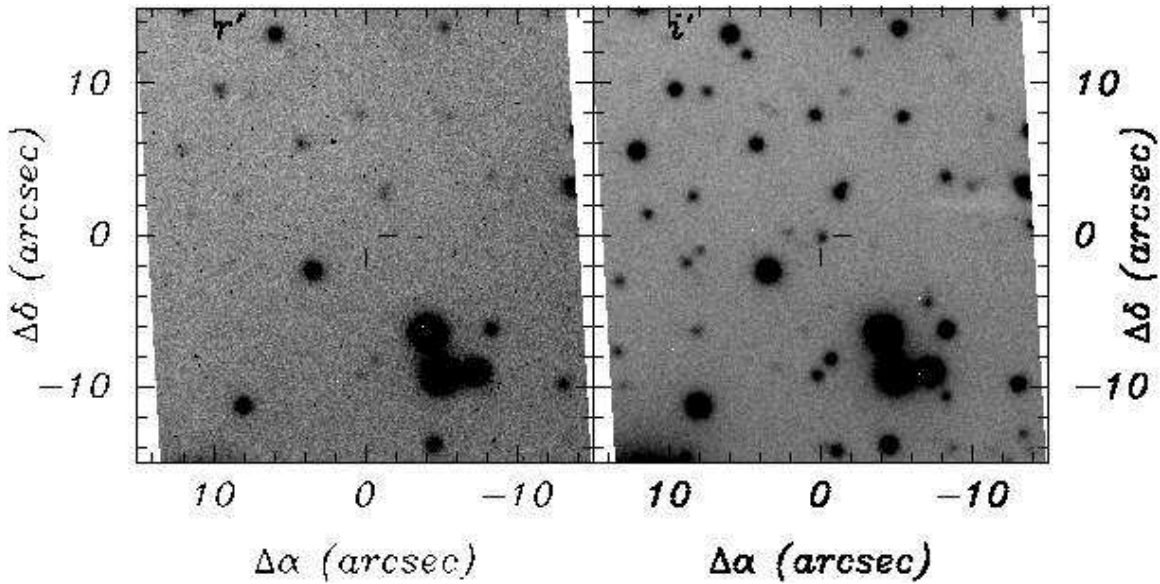}
\caption{MagIC $r^\prime$ (left) and $i^\prime$ (right) images of \sct.  In
  each case we plot a $30\arcsec$ box, with North up and East to the
  left.  The position of \mass\ is indicated by the tick marks.}
\label{fig:magic}
\end{figure*}

\subsection{Optical Photometry}
We performed photometric observations of \sct\ on 2006~August~28 with
the Magellan Instant Camera (MagIC) at an f/11 Nasmyth focus of the
6.5~m Baade (Magellan~I) telescope at Las Campanas Observatory in
Chile.  MagIC is a 2048$\times$2048 SITe CCD with a
0\farcs069~pixel$^{-1}$ plate scale and a 142\arcsec\ field of view.
Exposures of 1830~s in the $r^\prime$ filter and 3630~s in the
$i^\prime$ filter were obtained.  The conditions were
near-photometric, with $0\farcs7$ seeing in $r^\prime$ and $0\farcs6$
seeing in $i^\prime$.  We reduced the data according to standard
procedures in \texttt{IRAF} by subtracting overscan regions, merging
the data from four amplifiers, and flatfielding the data with twilight
flats.  We astrometrically calibrated the data by measuring the
positions of 50 2MASS stars and fitting for the transformation using
\texttt{ccmap}: the fit was characterized by an rms error of
$0\farcs07$ in each coordinate.

For purposes of photometry, we performed five 3-s observations of the
L110-232 standard field \citep{l92,s00} in each filter.  We
transformed the tabulated Kron-Cousins $R$ and $I$ magnitudes for 12
of the stars in that field using the results of \citet{fig+96} to
$r^\prime$ and $i^\prime$ magnitudes and determined zeropoints for the
observations.  These zeropoints agreed with the nominal values for
MagIC\footnote{See
\url{http://occult.mit.edu/instrumentation/magic/\#rpt\_txt}.} to
within our precision ($\approx 0.05$~mag).

As seen in Figure~\ref{fig:magic}, we detect an object at the position
of \mass\ in the $i^\prime$ image but not in the $r^\prime$ image.  We
estimate that the object has $r^\prime>25.2$ (3$\,\sigma$ limit) and
$i^\prime=23.64\pm0.15$, and we give the corresponding $i^\prime$-band
flux in Table~\ref{tab:ir}.  The photometry is consistent with the
non-detection of \mass\ in the DENIS $i$-band.

\subsection{Near-IR Spectroscopy}

\begin{figure*}
\plottwo{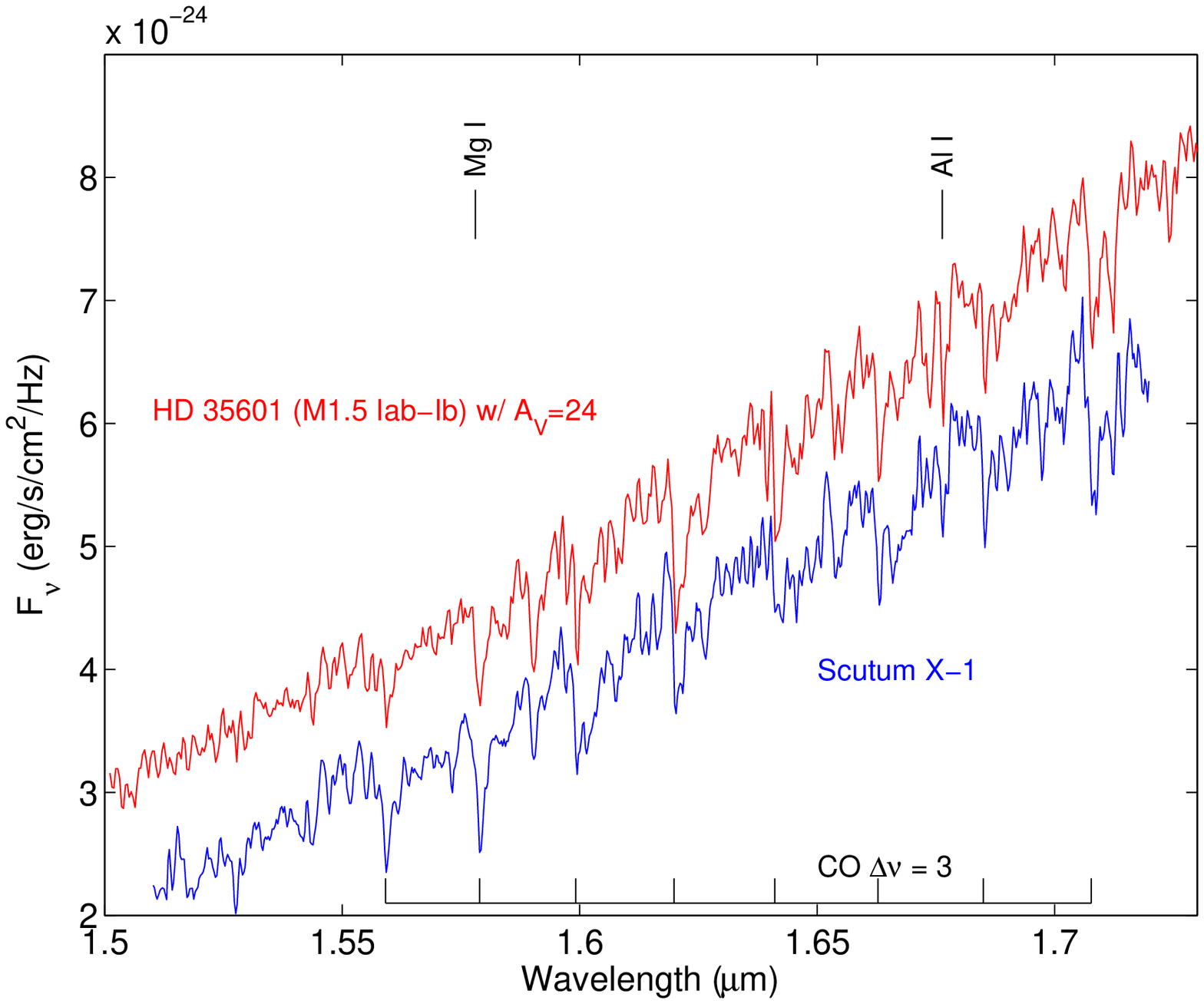}{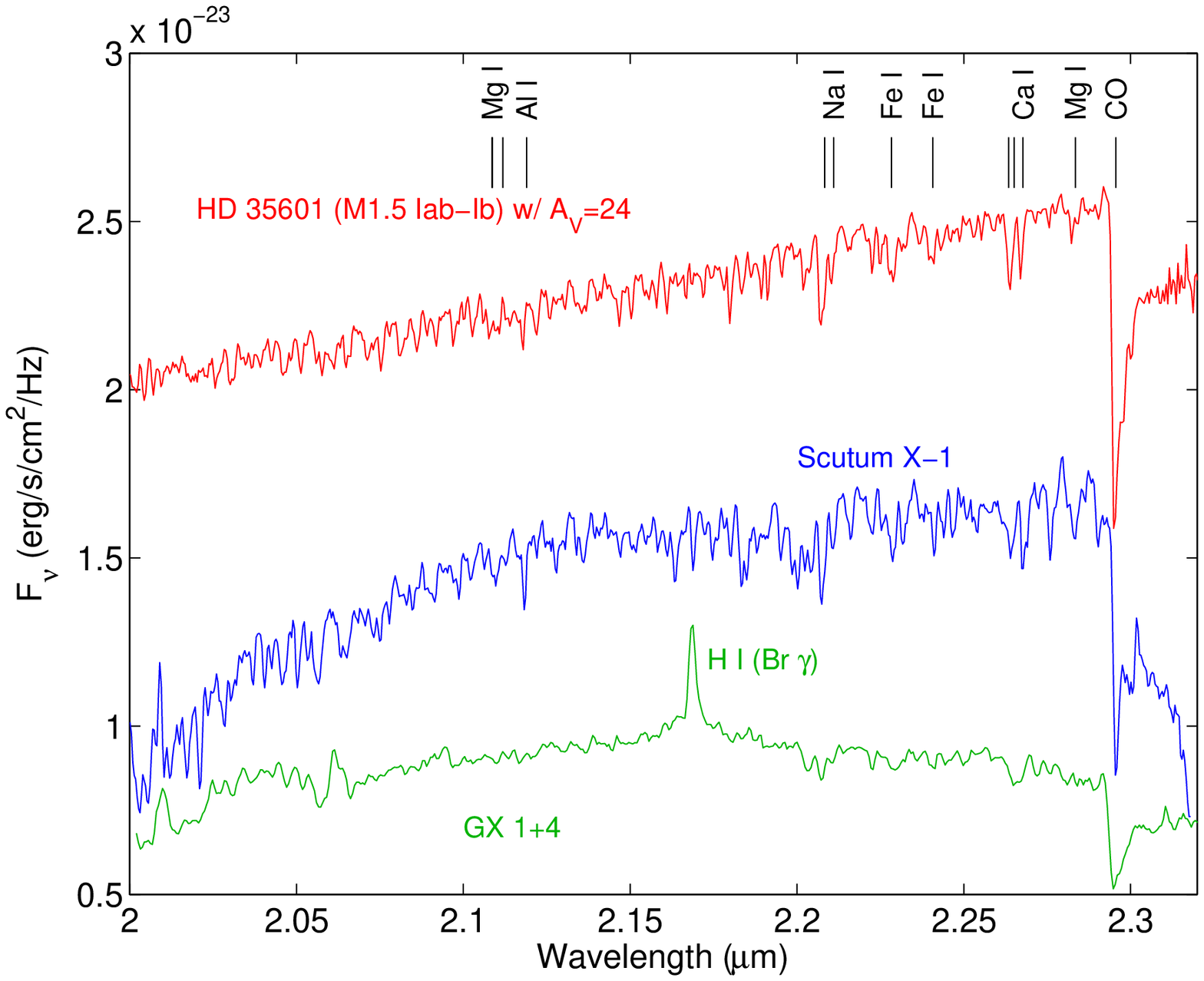}
\caption{Near-IR spectra of \mass\ in the $H$ (left) and $K$ (right)
  bands.  We show the spectrum of \sct\ along with the comparison
  stars HD~35601 (spectral type M1.5~Iab--Ib with an added extinction
  of $A_V=24$; from Rayner~et~al.\ 2006, in prep.)  and the X-ray
  binary GX~1+4 ($K$-band only; \citealt{cvkl98}) as labeled; the
  fluxes of both comparison stars have been shifted arbitrarily.  The
  main absorption lines in the $H$-band are the $\Delta \nu=3$ CO
  bands \citep{goorvitch94}, along with \ion{Mg}{1} and \ion{Al}{1}.
  In the $K$-band, the main lines are \ion{Mg}{1}, \ion{Al}{1},
  \ion{Na}{1}, \ion{Fe}{1}, and \ion{Ca}{1}, although GX~1+4 also
  shows a strong, broad \ion{H}{1} Br~$\gamma$ emission line; the very
  sharp drop at 2.29~$\mu$m is the beginning of the CO
  $\nu=2\rightarrow0$ band.  The flux scale is approximate.}
\label{fig:irspec}
\end{figure*}

To help determine the spectral type of the counterpart to \sct, we
undertook some limited near-IR spectroscopy.  The spectra were
obtained with the near-IR spectrograph NIRSPEC \citep{mbb+98} on the
Keck~II telescope in low-resolution ($R\sim1400$) mode and reduced
using the standard procedures described by \citet{ess+03}.  The
observations consisted of $4\times1$~s exposures in the $K$-band
(using the NIRSPEC-6 filter), and $4\times2$~s exposures in the
$H$-band (using the NIRSPEC-5 filter), on the night of
2006~October~3.  The resolution was $15$~\AA\ in $K$-band and
$10$~\AA\ in $H$-band.  We could not obtain standard star data from
the same night as the observations, but we did rough flux
calibrations of the $H$-band data using an observation from
2004~September of the A0V star HD~40335 and of the $K$-band data using an
observation from 2006~June of the A2V star HD~201941.  There were
variable clouds during the observations, and the calibration data are
not of the highest quality.  Therefore the absolute flux scale is
uncertain to about a factor of 2, and some smaller-scale deviations in
the continuum level are also present due to imperfect correction for
atmospheric transmission.

We show the spectra in Figure~\ref{fig:irspec}.  We can immediately
say that there are no strong emission lines such as one might expect
from the stellar wind in an X-ray binary.  Moreover, we do not see
strong \ion{H}{1} Paschen or Brackett series absorption that would indicate an early-type star
--- the type of star that is most commonly associated with X-ray
binaries of this pulse period (e.g., \citealt{bcc+97,cmpt99}).
Instead, we see strong CO absorption and lines from a number of
metals.  Comparing the lines that we identify with the sequences
presented in \citet{wh97}, \citet{mehs98}, and Rayner~et~al.\ (2006,
in prep.\footnote{See
\url{http://irtfweb.ifa.hawaii.edu/{\til}spex/spexlibrary/IRTFlibrary.html}.}),
we find the closest match is with late-K to early-M stars.  Given the
poor flux calibration we were not able to make a quantitative
classification, but the identification of the strong CO bands as well
as absorption from neutral metals seems secure; in particular, the
comparable strength of the \ion{Na}{1} doublet and the \ion{Ca}{1}
triplet along with the presence of \ion{Al}{1} in the $K$-band data
seem to indicate that the type is not too late.  We have not attempted
to determine the luminosity class of \sct, as the detailed
measurements necessary for that are beyond the tolerance of our
calibration, although the strength of the CO bands argues for
luminosity class I--III (see \citealt{mehs98}).

For comparison, in Figure~\ref{fig:irspec} we also plot the spectrum
of the M1.5~Iab--Ib star HD~35601 (taken from Rayner~et~al.\ 2006),
reddened with $A_V=24$ (see \S~\ref{sec:oir}).  A visual inspection
shows that our choice of comparison star is reasonable.  In the
$H$-band the match is particularly good, with the overall slope also
agreeing.  This then tells us that $A_V\approx 24$, and although we
have not corrected for any intrinsic reddening of the comparison star,
the intrinsic reddening should only be $\approx 1.5$~mag based on its
membership in the Aur~OB1 association (see \citealt{lmo+05}).  In the
$K$-band the match is not as good, with some curvature present in the
NIRSPEC data from poor calibration, but the absorption features and
the overall depth of the CO bands agree reasonably well. It is
possible, though, that the continuum shape in the $K$-band is a result
of water absorption in a very late M (later than M7 or so) star rather
than poor calibration (see \citealt*{crv05}).

\section{Discussion}
\label{sec:disc}

The detection of pulsations at approximately the same period as that
found by \citet{kktt91}, along with the position coincidence between
\axj\ and an alternate \heao\ diamond (Fig.~\ref{fig:image}; it is
also more or less consistent with the other X-ray positions), secures
the identification of \axj\ with \sct.  We now discuss the
implications of our measurements in more detail.

\subsection{Spin-Period Evolution}

\begin{figure}
\psfig{file=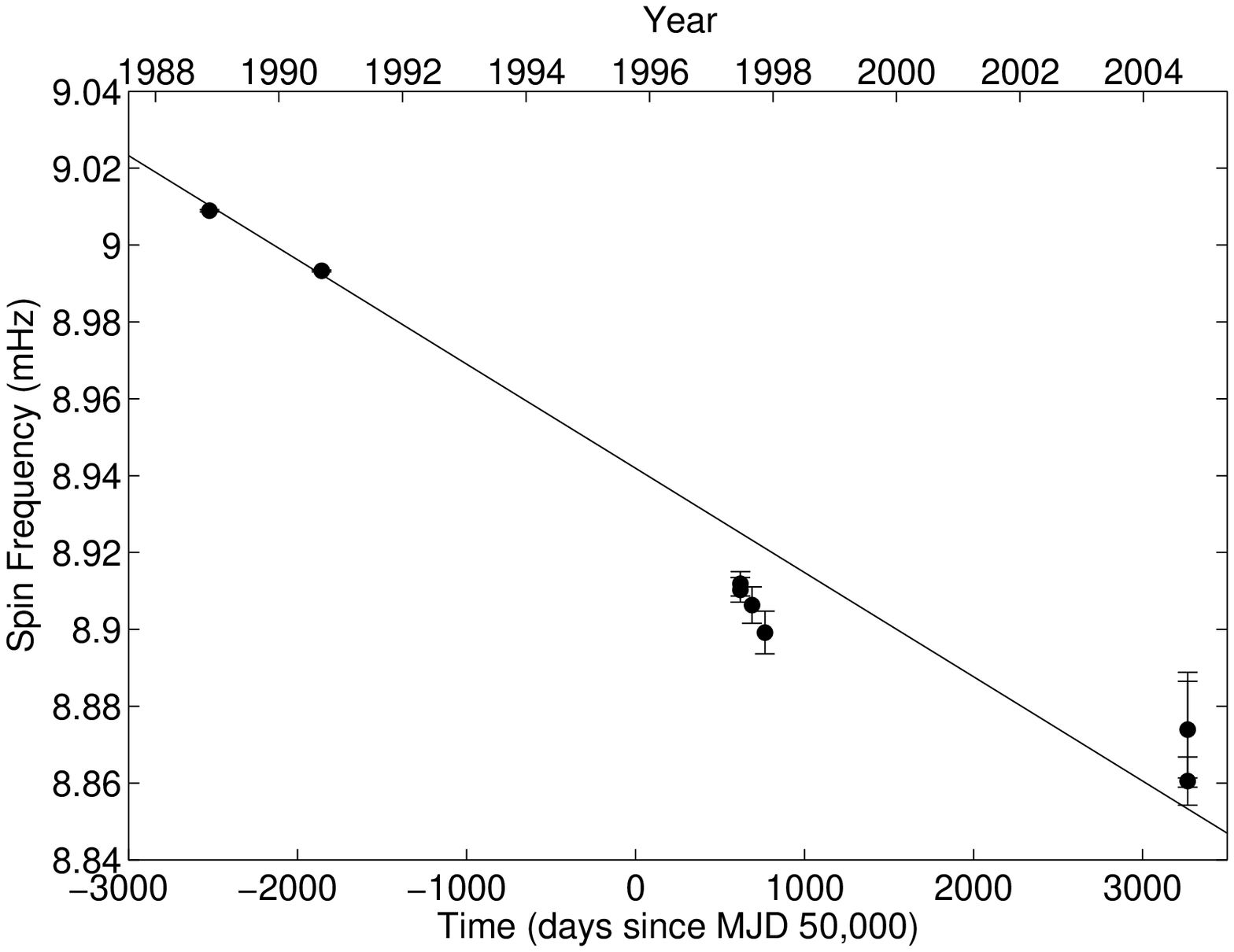,width=0.48\textwidth,clip=}
\caption{Spin-down of \sct, from \ginga\ \citep{yk93}, \rxte\ and
  \xmm\ data from Table~\ref{tab:per} (as labeled).  The line is a
  weighted fit with $\dot \nu=\expnt{-3.1}{-13}\mbox{ Hz s}^{-1}$.  }
\label{fig:pdot}
\end{figure}

Comparing the spin periods that we measure here to the values measured
by \citet{yk93} with the \textit{Ginga} satellite
(Table~\ref{tab:per}), we see evidence for secular spin-down over 17
years.  \citet{yk93} inferred a spin-down rate of $\dot
P=\expnt{3.3}{-9}\mbox{ s s}^{-1}$, although this was only based on
two measurements and could not account for possible torque variations
or periodic (i.e.\ orbital) changes.  Our period measurement fits this
trend reasonably well (Fig.~\ref{fig:pdot}): a weighted fit (which is
dominated by the \textit{Ginga} measurements) gives $\dot
\nu=\expnt{-3.1}{-13}\mbox{ Hz s}^{-1}$ ($\dot P=\expnt{3.9}{-9}\mbox{
s s}^{-1}$).  So it appears that the spin-down has largely been
steady, averaged over 17~years, although there certainly could have
been some small-scale torque variations visible in the departure of
the \rxte\ period measurements from the long-term trend.  This excludes the
possibility that the period change in \sct\ was due to orbital motion 
as considered by \citet{yk93}, as no companion can produce a Doppler
shift of 2~s over 17~yrs, and instead suggests that the spin-down
is due to external torques.  Overall, this scenario agrees with the
conclusion of \citet{yk93} that the compact object in \sct\ is a
neutron star rather than a white dwarf (cf.\ \citealt{ldomh+06}),
which would need a much larger torque than those that have been
observed to produce the measured $\dot P$.

\subsection{X-ray Spectrum}
Even if we ignore the presence or absence of the Fe line, 
%which could easily be due to background-subtraction problems, 
our best-fit spectral parameters are not the same as those estimated
by \citet{kktt91}, who found a slightly softer power-law ($\Gamma=2.0$
vs.\ $1.5$ here) and more absorption ($N_{\rm
H}=\expnt{(2-4)}{23}\mbox{ cm}^{-2}$ vs.\ $\expnt{8}{22}\mbox{
cm}^{-2}$ here).  In fact, most of the X-ray observations to date
(which have not been of the same quality as the \xmm\ observation)
have inferred column densities above $10^{23}\mbox{ cm}^{-2}$.  While
this could just be an effect of instrumental cross-calibration and
inconsistent fitting techniques, the difference is quite large.  This
could be due to variations in absorption by matter associated with the
\sct\ system over the $17$~yrs between the observations (see
\citealt{yk93}); in particular, variations in $N_{\rm H}$ are often
associated with absorption from a variable stellar wind
\citep[e.g.,][]{ws84,cr97}, and a decrease in $N_{\rm H}$ can even be
correlated with a decrease in the Fe equivalent width.  Overall, the
flux has decreased by about a factor of 4 from the \textit{Ginga}
observations to the \xmm\ observations (we find a flux of $\approx
0.4$~mCrab), and the variability of the pulsed amplitude during the
\rxte\ observations over the course of several months
(Tab.~\ref{tab:xte}) may be as much as a factor of 10.  The unabsorbed
flux in the \xmm\ observations implies a luminosity $L_{\rm
X}=\expnt{1.4}{33}d_{\rm kpc}^2\mbox{ erg s}^{-1}$ in the 0.5--10~keV
band, where the distance to \sct\ is $d_{\rm kpc}$ kiloparsecs.

\subsection{Optical/IR Counterpart: Constraints on Extinction and Distance}
\label{sec:oir}
The very red colors of \mass\ imply a large extinction, as no stars
have intrinsic colors nearly that red.  The reddest main-sequence star
listed in \citet{allen} has $J-K_{\rm s}\approx 1$ (M5V), while the
reddest giant star has $J-K_{\rm s}\approx 1.2$ (M5III).  If we assume
intrinsic colors of $J-K_{\rm s} \leq 1.2$ for \mass\ and that the
emission we see is photospheric (no excess from a disk or wind), this
then implies $A_V \gsim 24$ (hence the giant track shown in
Figure~\ref{fig:cmd}).  Such extinction makes \mass\ very faint in the
optical, which is consistent with our measured $i^\prime$ magnitude
and with the upper limits at $r^\prime$ and in other bluer bands.
This extinction agrees with the slope of the $H$-band continuum,
although given our calibration uncertainties that is not a strong
statement.  The X-ray absorption implied by this extinction, using the
relation of \citet{ps95}, is $N_{\rm H} \gsim \expnt{4}{22}\mbox{
cm}^{-2}$, which is consistent with our spectroscopic result.

The Galactic extinction model of \citet*{dcllc03} predicts that
$A_V=19$ at 7.5~kpc, and $A_V=28$ at 10~kpc, although at these
distances and extinctions, the model is not very well constrained.  So
based on extinction alone, and assuming that the extinction is
extrinsic to the sources, we would estimate a distance of $\gsim
8$~kpc for \sct, which is consistent with the assertion of
\citet{kkk+90}, that \sct\ is likely in a spiral arm at $\approx
10$~kpc.

However, while we know that \mass\ is heavily reddened, with the
limited data that we have it is difficult to determine its intrinsic
colors and stellar type with any precision.  A rough fit to the
$i^\prime JHK_{\rm s}$ photometry (and the $r^{\prime}$ upper limit)
is reasonably consistent with a late-type star at $A_V\lsim 30$.  This
inference is consistent with what we deduce from the near-IR
spectroscopy.  In general, for a given stellar type the distance to
\mass\ is
\[
\log_{10} d_{\rm kpc}=0.2\left[(K_{\rm s,obs}-0.11 A_{V}) + (V-K)_0-M_V\right]-2,
\]
where $K_{\rm s,obs}=6.55$, $A_{K}/A_V=0.11$, and $(V-K)_0$ and
$M_V$ are the color and absolute magnitude of the star, which we take
from \citet{allen}.  We determine $A_V$ by
\[
A_V=\frac{(J-K_{\rm s})_{\rm obs}-(J-K)_0}{(0.29-0.11)},
\]
where we observe $(J-K_{\rm s})_{\rm obs}=5.51$, $(J-K)_0$ is the
intrinsic color of the star, and $A_J/A_V=0.29$.  If we take \mass\ to
be a late-type supergiant, we have (for M0I) $(V-K)_0\approx3.80$,
$(J-K)_0\approx 0.9$, and $M_V\approx-5.6$, so we find a distance of
$\approx 4$~kpc.  This is slightly smaller than that implied by the
extinction, but gives a reasonable X-ray luminosity of $\approx
\expnt{2}{34}\mbox{ erg s}^{-1}$.

%% The assumed colors and magnitudes used above are strictly appropriate
%% for stars only and ignore excesses.  In
%% particular, \citet{cegm+05} studied a number of systems in the Small
%% Magellanic Cloud, and based on this the absolute $V$ magnitudes are
%% between $-3$ and $-5$.  The observed $V-K$ color varies from what we
%% have assumed due to infrared excesses, and values from 0.2 to 1.0 are
%% typical, so $\Delta K\approx0-2$ \citep{dwb+94}.  These revised values
%% do not change the distance too much, and can only increase it to $\sim
%% 700$~pc (meaning $L_{\rm X}\sim \expnt{7}{32}\mbox{ erg s}^{-1}$) at
%% the extreme ranges of these values.  

%% Even allowing for some infrared excess, this distance is far too small
%% to accommodate the X-ray luminosity and optical extinction.  We must
%% instead take \mass\ to be an OB supergiant.  With this model, we get
%% considerably larger distances of $\gsim 1$~kpc that are more
%% compatible with the X-ray and optical/IR data, although the precise
%% distance is difficult to determine as the optical luminosities of such
%% stars vary widely \citep{lzdw+98}.

\subsection{Comparison with GX~1+4}
Given its infrared and X-ray characteristics, \sct\ resembles the
2-minute X-ray pulsar GX~1+4, which is a symbiotic system with an
M-giant mass donor (\citealt*{dmb77}; \citealt*{cr97,cvkl98}).  GX~1+4
generally is quite similar to \sct: it has a similar spin-period and
showed a prolonged period of relatively faint ($<0.5$--2~mCrab)
emission with steady spin-down \citep{cbf+97}, although it had
previous bright periods associated with spin-up.  We note, though,
that we did not see the strong \ion{H}{1} emission lines one would
expect from a symbiotic star (Fig.~\ref{fig:irspec}), but this may be
a result of variability or viewing geometry (see, e.g.,
\citealt{mop+06}).  We also note that GX~1+4 appears somewhat less
luminous in the near-IR: GX~1+4 has $K\approx 8.1$ with $A_V\approx
5$, which imply $M_{K}\approx-5.6$ or $M_V\approx -0.5$
\citep{hfj+06}.  If the \sct\ system contains a star like this it
would only be 500~pc distant; the observed extinction in the
optical/near-IR and low energy X-ray absorption would be hard to
explain, and the X-ray luminosity would be very low.

\section{Conclusions}
\label{sec:conc}
It is possible  that \sct\ is a nearby low-mass X-ray binary with
an evolved giant companion.  This would then imply a low X-ray
luminosity --- comparable to those implied by \citet{mop+06} --- and
that the X-ray and optical/IR absorption are intrinsic to the source,
perhaps caused by material in the stellar wind (e.g.,
\citealt{rsgs03,fc04}).  This would be consistent with the possible
variations in $N_{\rm H}$ seen between our \xmm\ data and previous
observations, but those variations are far from robustly determined.
Additionally, the reasonable agreement between the X-ray and
optical/IR column densities \citep[c.f.][]{fc04,kmr06} suggests that the
absorbing material is largely distributed along the line of sight.

We  therefore believe it  very likely that \sct\  is an X-ray binary
with a giant or supergiant late-type companion located at a distance
of $\gsim 4$~kpc.  In this case, the mass donor in \sct\ would be
larger than several hundred $R_{\odot}$ (the companion to GX~1+4 has a
radius $\approx 100R_{\odot}$) and significantly more massive than in
the previous scenario (i.e., a supergiant of $\approx 10M_\odot$
rather than an evolved giant of 1--2$M_{\odot}$).  With such a large
companion, even at a distance of 10~kpc the X-ray luminosity would be
very low for steady Roche-lobe overflow accretion, so therefore the
accretion must be either wind-fed or in an elliptical orbit.  With
either companion (a giant or supergiant), if we assume that the
accretion is wind-fed, we can set a lower limit in the orbital period
$P_{\rm orb}$.  For a circular orbit and M0I star (with stellar radius
$R_*=500R_{\odot}$ and mass $M_*=13M_{\odot}$), we would need $P_{\rm
orb} \gsim 2$~yr so that it was contained within its Roche lobe, while
a star like the companion to GX~1+4 ($R_*=100R_{\odot}$ and
$M_*=1.2M_{\odot}$) would have $P_{\rm orb} \gsim 0.8$~yr.
% if this really is high mass, can it be used to set limits on NS progenitor?

With a large companion, \sct\ is a good candidate for an eclipsing
system since the companion will occult the neutron star with a
significant probability \citep{rc02}, and such eclipsing systems offer
valuable constraints on the neutron star mass distribution
\citep*{vkvpz95}.  Further X-ray observations to measure the orbit of
\sct\ and fully study the spectrum, along with optical/IR observations
to determine the parameters of \mass, should be quite valuable in
unraveling the nature of the system.

\acknowledgements We thank A.~Burgasser for conducting the Magellan
observations and helpful discussion, and  C.~Steidel
for assistance with the near-IR spectroscopy.  The Digitized Sky
Surveys were produced at the Space Telescope Science Institute under
U.S.\ Government grant NAG~W-2166. The images of these surveys are
based on photographic data obtained using the Oschin Schmidt Telescope
on Palomar Mountain and the UK Schmidt Telescope. The plates were
processed into the present compressed digital form with the permission
of these institutions.  This publication makes use of data products
from the Two Micron All Sky Survey, which is a joint project of the
University of Massachusetts and the Infrared Processing and Analysis
Center/California Institute of Technology, funded by the National
Aeronautics and Space Administration and the National Science
Foundation.  This research made use of data products from the
Midcourse Space Experiment. Processing of the data was funded by the
Ballistic Missile Defense Organization with additional support from
NASA Office of Space Science. This research has also made use of the
NASA/ IPAC Infrared Science Archive, which is operated by the Jet
Propulsion Laboratory, California Institute of Technology, under
contract with the National Aeronautics and Space Administration.  This
work is based in part on observations made with the Spitzer Space
Telescope, which is operated by the Jet Propulsion Laboratory,
California Institute of Technology under a contract with NASA.  The
DENIS project has been partly funded by the SCIENCE and the HCM plans
of the European Commission under grants CT920791 and CT940627.  It is
supported by INSU, MEN and CNRS in France, by the State of
Baden-W\"urttemberg in Germany, by DGICYT in Spain, by CNR in Italy,
by FFwFBWF in Austria, by FAPESP in Brazil, by OTKA grants F-4239 and
F-013990 in Hungary, and by the ESO C\&EE grant A-04-046.  This
research has made use of the SIMBAD database, operated at CDS,
Strasbourg, France.  This research has made use of data obtained from
the High Energy Astrophysics Science Archive Research Center
(HEASARC), provided by NASA's Goddard Space Flight Center.  This
research has made use of SAOImage DS9, developed by the Smithsonian
Astrophysical Observatory.  Finally, we wish to extend special thanks
to those of Hawaiian ancestry on whose sacred mountain we are
privileged to be guests. Without their generous hospitality, some of
the observations presented herein would not have been possible.

{\it Facilities:} \facility{Magellan:Baade (MagIC)}, \facility{XMM
  (EPIC)}, \facility{RXTE (PCA)}

%\bibliography{xray,myrefs,xrb,magrefs,casA}

%% --------------------------------------------------------------------
%% Thu Dec 21 14:25:28 2006
%%   This file was generated automagically from the files
%%   ms.bbl and ms.tex using
%%     /Users/dlk//perl/nat2jour.pl
%%   This file should accompany ms-aas.tex.
%% --------------------------------------------------------------------

\end{document}